\documentclass[11pt]{elsart}
\usepackage[dvips]{graphicx}
\usepackage{amsmath}
\usepackage{array}
\usepackage{amssymb}
\usepackage{amsmath}
\usepackage{hhline}
\usepackage{longtable}

\begin{document}
\begin{frontmatter}

\title{Fractal scale-free networks resistant to disease spread}

\author[lable1,label2]{Zhongzhi Zhang\corauthref{zzz}}
\corauth[zzz]{Corresponding author.} \ead{zhangzz@fudan.edu.cn}
\author[lable1,label2]{Shuigeng Zhou}
\ead{sgzhou@fudan.edu.cn}
\author[lable1,label2]{Tao Zou}
\ead{zoutao@fudan.edu.cn}
\author[label3]{Guisheng Chen}

\address[lable1]{Department of Computer Science and Engineering, Fudan
University, Shanghai 200433, China}
\address[label2]{Shanghai Key Lab of Intelligent Information Processing,
Fudan University, Shanghai 200433, China}
\address[label3]{China Institute of Electronic System Engineering, Beijing 100840, China}

\begin{abstract}
In contrast to the conventional wisdom that scale-free networks are
prone to epidemic propagation, in the paper we present that disease
spreading is inhibited in fractal scale-free networks. We first
propose a novel network model and show that it simultaneously has
the following rich topological properties: scale-free degree
distribution, tunable clustering coefficient, ``large-world"
behavior, and fractal scaling. Existing network models do not
display these characteristics. Then, we investigate the
susceptible-infected-removed (SIR) model of the propagation of
diseases in our fractal scale-free networks by mapping it to bond
percolation process. We find an existence of nonzero tunable
epidemic thresholds by making use of the renormalization group
technique, which implies that power-law degree distribution does not
suffice to characterize the epidemic dynamics on top of scale-free
networks. We argue that the epidemic dynamics are determined by the
topological properties, especially the fractality and its
accompanying ``large-world" behavior.
\begin{keyword}
Complex networks\sep Disease spread\sep Fractal networks\sep
Scale-free networks
\PACS 89.75.-k \sep 64.60.Ak \sep 87.23.Ge
\end{keyword}
\end{abstract}
\date{\today}
\end{frontmatter}



\section{Introduction}

In recent years, there has been much interest in the study of the
structure and dynamics of complex
networks~\cite{AlBa02,DoMe02,Ne03,BoLaMoChHw06}. One aspect that has
received considerable attention is the epidemic spreading taking
place on top of networks~\cite{BaBaPaVe05}, which is relevant to
computer virus diffusion, information and rumor spreading, and so
on. In the study of epidemic spreading, the notion of thresholds is
a crucial problem since it finds an intermediate practical
application in disease eradication and vaccination
programs~\cite{Ba75,He00}. In homogeneous networks, there is an
existence of nonzero infection threshold, if the spreading rate is
above the threshold, the infection spreads and becomes endemic,
otherwise the infection dies outs quickly. However, recent studies
demonstrate that  the threshold is absent in heterogeneous
scale-free networks~\cite{BaAl99,PaVe01a,PaVe01b,MoPaVe02}. Thus, it
is important to identify what characteristics of network structure
determine the presence or not of  epidemic thresholds.

To date the influences of most structural properties on disease
dynamics have been studied, which include degree
distribution~\cite{PaVe01a,PaVe01b,MoPaVe02}, clustering
coefficient~\cite{SeBo06}, and degree correlations~\cite{BoPaVe03}.
However, these features do not suffice to characterize the
architecture of a network~\cite{CoRoTrVi07}. Very recently, by
introducing and applying box-covering (renormalization) technique,
Song, Havlin and Makse found the presence of fractal scaling in a
variety of real networks~\cite{SoHaMa05,SoHaMa06}. Examples of
fractal networks include the WWW, actor collaboration network,
metabolic network, and yeast protein interaction network
~\cite{YoRaOr06}. The fractal topology is often characterized
through two quantities: fractal dimension $d_B$ and degree exponent
of the boxes $d_k$, both of which can be calculated by the
box-counting algorithm~\cite{SoGaHaMa07,KiGoKaKi07}. The scaling of
the minimum possible number of boxes $N_{B}$ of linear size
$\ell_{B}$ required to cover the network defines the fractal
dimension $d_B$, namely $N_{B} \thicksim \ell_{B}^{-d_{B}}$.
Similarly, the degree exponent of the boxes $d_k$ can be found via
$k_{B}(\ell_{B})/ k_{hub} \thicksim \ell_{B}^{-d_{k}}$, where
$k_{B}(\ell_{B})$ is the degree of a box in the renormalized
network, and $k_{hub}$ the degree of the most-connected node inside
the corresponding box. Interestingly, for fractal scale-free
networks with degree distribution $P(k)\sim k^{-\gamma}$, the two
exponents, $d_{B}$ and $d_{k}$, are related to each other through
the following universal relation: $\gamma
=1+d_{B}/d_{k}$~\cite{SoHaMa05}.

Fractality is now acknowledged as a fundamental property of a
complex network~\cite{CoRoTrVi07}. It relates to a lot of aspects of
network structure and dynamics running on the network. For example,
in fractal networks the correlation between degree and betweenness
centrality of nodes is much weaker than that in non-fractal
networks~\cite{KiHaPaRiPaSt07}. In addition, several studies
uncovered that fractal networks are not
assortative~\cite{SoHaMa06,YoRaOr06,ZhZhZo07}. The peculiar
structural nature of fractal networks make them exhibit distinct
dynamics. It is known that fractal scale-free networks are more
robust than non-fractal ones against malicious attacks on hub
nodes~\cite{SoHaMa06,ZhZhZo07}. On the other hand, fractal networks
and their non-fractal counterparts also display disparate phenomena
of other dynamics, such as cooperation~\cite{HiBe06,Hi07},
synchronization~\cite{ZhZhZo07}, transport~\cite{GaSoHaMa07}, and
first-passage time~\cite{RoHaAv07,CoBeTeVoKl07}. Despite of the
ubiquity of fractal feature and its important impacts on dynamical
processes, the dynamics of disease outbreaks in fractal networks has
been far less investigated.

In this current paper, we focus on the effects of fractality on the
dynamics of disease in fractal scale-free networks. Firstly, we
propose an algorithm to create a class of fractal scale-free graphs
by introducing a control parameter $q$. Secondly, we give in detail
a scrutiny of the network architecture. The analysis results show
that this class of networks have unique topologies. They are
simultaneously scale-free, fractal, `large-world', and have tunable
clustering coefficient. Thirdly, we study a paradigmatic
epidemiological model~\cite{Ba75,He00}, namely the
susceptible-infected-removed (SIR) model on the proposed fractal
graphs. By mapping the SIR model to a bond percolation problem and
using the renormalization-group theory, we find the existence of
non-zero epidemic thresholds as a function of $q$. We also provide
an explanation for our findings.

\section{Network construction and topological properties}

This section is devoted to the construction and the relevant
structural properties of the networks under consideration, such as
degree distribution, clustering coefficient, average path length
(APL), and fractality.

\begin{figure}[h]
\begin{center}
\includegraphics[width=0.6\textwidth]{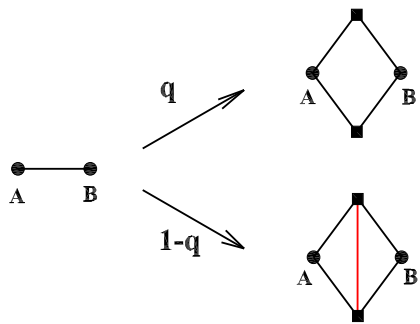}
\caption{(Color online) Iterative construction method of the fractal
networks. Each iterative link is replaced by a connected cluster on
the right-hand side of the arrow. The red link is a noniterated
link.}\label{fig1}
\end{center}
\end{figure}

\subsection{Construction algorithm}

\begin{figure}[h]
\begin{center}
\includegraphics[width=.7\linewidth,trim=100 0 140 20]{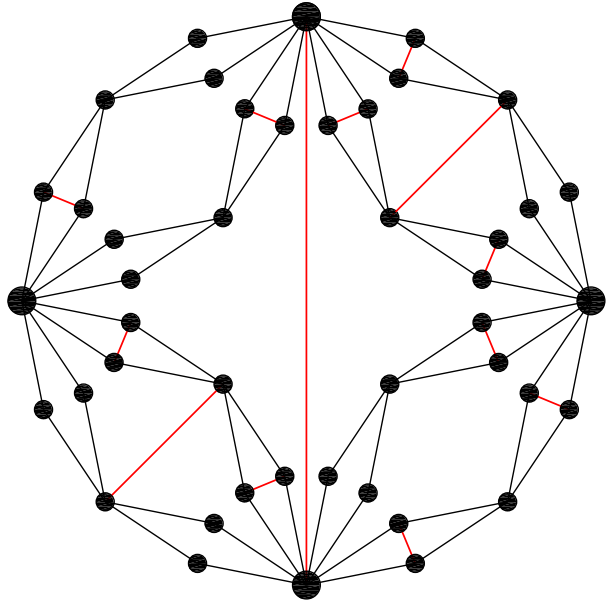}
\caption{(Color online) Sketch of a network after three iterations
for the particular case of $q=0.6$.}\label{fig2}
\end{center}
\end{figure}

The proposed fractal networks have two categories of bonds (links or
edges): iterative bonds and noniterated bonds which are depicted as
solid and dashed lines, respectively. The networks are constructed
in an iterative way as shown in Fig.~\ref{fig1}. Let $F_{t}$ ($t\geq
0$) denote the networks after $t$ iterations. Then the networks are
built in the following way: For $t=0$, $F_{0}$ is two nodes
(vertices) connected by an iterative edge. For $t\geq 1$, $F_{t}$ is
obtained from $F_{t-1}$. We replace each existing iterative bond in
$F_{t-1}$ either by a connected cluster of links on the top middle
of Fig.~\ref{fig1} with probability $q$, or by the connected cluster
on the bottom right with complementary probability $1-q$. The
growing process is repeated $t$ times, with the fractal graphs
obtained in the limit $t \to \infty$. Figure~\ref{fig2} shows a
network after three-generation growth for a specific case of
$q=0.6$.

Next we compute the numbers of total nodes and edges in $F_t$. Let
$L_v(t)$, $L_i(t)$ and $L_n(t)$ be the numbers of nodes, iterative
edges, and noniterated edges created at step $t$, respectively. Note
that each of the existing iterative edges yields two nodes and four
new iterative edges; at the same time this original iterative edge
itself is deleted, which means that $L_i(t)$ is also the total
number of iterative edges at time $t$. Then we have
$L_v(t)=2\,L_i(t-1)$ and $L_i(t)=4\,L_i(t-1)$ for all $t>0$.
Considering the initial condition $L_i(0)=1$, one can obtain
$L_i(t)=4^t$ and $L_v(t)=2\times 4^{t-1}$. Thus the number of total
nodes $N_t$ present at step $t$ is
\begin{eqnarray}\label{Nt1}
N_t=\sum_{t_i=0}^{t}L_v(t_i)=\frac{2\times4^{t}+4}{3}.
\end{eqnarray}

On the other hand, at each construction step, each of the existing
iterative edges may yield one noniterated link with probability
$1-q$, so the expected value of $L_n(t)$ is $(1-q)\,L_i(t-1)$ for
$t\geq 1$, i.e., $L_n(t)=(1-q)\, 4^{t-1}$. Therefore, the total
number of edges $E_t$ present at step $t$ is
\begin{eqnarray}\label{Et1}
E_t=L_i(t)+\sum_{t_i=1}^{t}L_n(t_i)=\frac{(4-q)\,4^{t}-(1-q)}{3}.
\end{eqnarray}
 The average node degree after $t$ iterations is $\langle k
 \rangle_t=\frac{2\,E_t}{N_t}$,
which approaches $4-q$ in the infinite $t$ limit.

\subsection{Degree distribution}
When a new node $u$ enters the system at step $t_u$ ($t_u\geq 1$),
it has two iterative edges. At the same time, with probability $1-q$
one noniterated edge is created and linked to node $u$. Let
$L_i(u,t)$ be the number of iterative links emanated from node $u$
at step $t$, then $L_i(u, t_u)=2$. Notice that at any subsequent
step each iterative edge of $u$ is broken and generates two new
iterative edges linked to $u$. Thus
$L_i(u,t)=2\,L_i(u,t-1)=2^{t-t_{u}+1}$.

We define $k_u(t)$ as the degree of node $u$ at time $t$, then we
have
\begin{equation}\label{ki}
\left\{\begin{array}{lc} {k_u(t)=2^{t-t_{u}+1},}& \ \hbox{with probability}\ q,\\
{{k_u(t)=2^{t-t_{u}+1}+1},}& \ \hbox{with probability}\
1-q,\end{array} \right.
\end{equation}
where the last term 1 in the second formula represents the
noniterated link connected to node $u$. For the initial two nodes
created at step 0, neither of them has a noniterated link, both
nodes have a degree of $2^{t}$.

Equation~(\ref{ki}) indicates that the degree spectrum of the
networks is not continuous. It follows that the cumulative degree
distribution~\cite{DoGoMe02,ZhZhZoChGu07} is given by $P_{\rm
cum}(k)=\frac{N_{t,k}}{N_t}$, where $N_{t,k}$ is the number of nodes
whose degree is not less than $k$. When $t$ is large enough, we find
$P_{\rm cum}(k)\approx k^{-2}$. So the degree distribution follows a
power-law form with the exponent $\gamma=3$, which is independent of
$q$. The same degree exponent has been obtained in the famous
Barab\'asi-Albert (BA) model~\cite{BaAl99}.

\subsection{Clustering Coefficient}
The clustering coefficient~\cite{WaSt98} of a node $u$ with degree
$k_u$ is the probability that a pair of neighbors of $u$ are
themselves connected, which is given by $C(k_u)=2b_u/[k_u( k_u-1)]$,
where $b_u$ is the number of existing connections between the $k_u$
neighbors of $u$. For our networks, the clustering coefficient
$C(k)$ for a single node with degree $k$ can be evaluated exactly.
Note that except for those nodes born at step $t$, all existing
links among the neighbors of a given node are noniterated ones,
whose number is ease to calculate. For the initial two nodes, the
expected existing noniterated links among the neighbors is
$(1-q)2^{t-1}$. For each of those nodes created at step $\phi$
$(0<\phi<t)$, there are average $(1-q)2^{t-\phi}$ noniterated links
among its neighbors. Finally, for the nodes generated at step $t$,
some of them have a degree of $3$, the number of links between the
neighbors of each is $2$; the others have a degree of $2$, their
clustering coefficient is zero. Thus, there is a one-to-one
correspondence between the clustering coefficient $C(k)$ of the node
and its degree $k$:
\begin{equation}\label{Ck}
C(k)=\left\{\begin{array}{lc} {\displaystyle{ (1-q) / (k-1) } }
& \ \hbox{for}\ k=2^m (2 \leq m\leq t),\\
{\displaystyle{(1-q) / k} }
& \ \hbox{for}\  k=2^m+1 (2 \leq m\leq t),\\
{\displaystyle{0} }
& \ \hbox{for}\  k=2,\\
{\displaystyle{ 2 / k } } & \ \hbox{for} \  k=3.\end{array} \right.
\end{equation}
\begin{figure}[h]
\centering\includegraphics[width=0.6\textwidth]{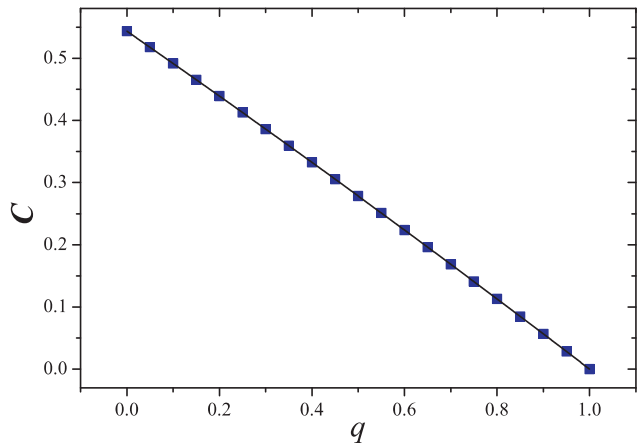}
\caption{(Color online) The average clustering coefficient $C$ of
the whole network as a function of $q$. It can be tunable
systematically by changing $q$. The squares are the simulation
results, and the line represents the analytical expression given by
Eq.~(\ref{ACC}). }\label{fig3}
\end{figure}
Using Eq.~(\ref{Ck}), we can obtain the clustering $\bar{C}_t$ of
whole the network at step $t$, which is defined as the average
clustering coefficient of all individual nodes. Then we have
\begin{eqnarray}\label{ACC}
\bar{C}_t= &\quad&
    \frac{1-q}{N_{t}} \Bigg[\frac{L_v(0)}{2^t-1}+ \sum_{r=1}^{t-1}\frac{qL_v(r)}{K_r-1} \nonumber\\  &+& \sum_{r=1}^{t-1}\frac{(1-q)L_v(r)}{K_r+1}+\frac{2\,L_v(t)}{K_t+1}\Bigg] ,
\end{eqnarray}
where $K_r=2^{t-r+1}$ and $K_t=2$. In the infinite network order
limit ($N_{t}\rightarrow \infty$), $\bar{C}_t$ converges to a
nonzero value $C$ as a function of parameter $q$. In the two extreme
cases of $q=0$ and $q=1$, $C$ are 0 and 0.5435~\cite{ZhZhZoGu08},
respectively. When $q$ increases from 0 to 1, $C$ grows from  0 to
0.5435, see Fig.~\ref{fig3}. Thus, the parameter $q$ in our model
introduces the clustering effect by allowing the formation of
triangles. Furthermore, the relation between $C$ and $q$ is almost
linear, as depicted in Fig.~\ref{fig3}.

\subsection{Average path length}

Shortest paths play an important role both in the transport and
communication within a network and in the characterization of the
internal structure of the
network~\cite{FrFrHo04,HoSiFrFrSu05,DoMeOl06,ZhChZhFaGuZo08}. Let
$d_{ij}$ represent the shortest path length from node $i$ to $j$,
then the average path length (APL) $d_{t}$ of $F_t$ is defined as
the mean of $d_{ij}$ over all couples of nodes in the network, and
the maximum value $D_t$ of $d_{ij}$ is called the diameter of the
network.

For general $q$, it is difficult to derive a closed formula for the
APL $d_{t}$ of network $F_t$. But for two limiting cases of $q=0$
and $q=1$, both the networks are deterministic ones, we can obtain
the analytic solutions for APL, denoted as $d_{t}^{0}$ and
$d_{t}^{1}$, respectively. In the large $t$ limit, $d_{t}^{0}\approx
\frac{8}{21}2^t$~\cite{ZhZhZoGu08} and $d_{t}^{1}\approx
\frac{11}{21}2^t$~\cite{HiBe06}. On the other hand, for large $t$
limit, $N_t \sim 4^t$, so both $d_{t}^{0}$ and $d_{t}^{1}$ grow as a
square power of the number of network nodes.

By construction, it is obvious that $d_{t}^{0}$ and $d_{t}^{1}$ are
the lower and upper bound of APL for network $F_t$, respectively. As
a matter of fact, if we denote the specific $q=1$ case of the
network as $F_t^1$ (it has no noniterated edges, and thus the
maximum APL), then one can obtain $F_t$ from $F_t^1$ by adding
noniterated edges in a certain way. The less the parameter $q$, the
more the added noniterated edges. In the case of $q=0$, the network
is the densest one with the minimum APL. Therefore, the APL is a
decreasing function of $q$. As $q$ drops from 1 to 0, $d_{t}$
increases from $\frac{8}{21}2^t$ at $q=1$ to $\frac{11}{21}2^t$ at
$q=0$. From above arguments, we can conclude that for the full range
of $q$ between 0 and 1, the APL $d_{t}$ has an exponential growth
with network size, which indicates that the considered  networks
$F_t$ are not small worlds.

\subsection{Fractal dimension}
To determine the fractal dimension, we follow the mathematical
framework introduced in Ref.~\cite{SoHaMa06}. We are concerned about
three quantities, network size $N_t$, network diameter $D_t$, and
degree $k_u(t)$ of a given node $u$. By construction, we can easily
see that in the infinite $t$ limit, these quantities grow obeying
the following relation: $N_t\simeq 4\,N_{t-1}$, $D_{t}=2\,D_{t-1}$,
$k_u(t)\simeq 2\,k_u(t-1)$. Thus, for large networks, $N_t$,
$k_u(t)$ and $D_t$ increase by a factor of $f_N=4$, $f_k=2$ and
$f_D=2$, respectively.

From above obtained microscopic parameters demonstrating the
mechanism for network growth, we can derive the scaling exponents:
the fractal dimension $d_B =\frac{\ln f_N}{\ln f_D}=2$ and the
degree exponent of boxes $d_k =\frac{\ln f_k}{\ln f_D}=1$. According
to the scaling relation of fractal scale-free networks, the exponent
of the degree distribution satisfies $\gamma =1+\frac{d_B}{d_k}=3$,
giving the same $\gamma$ as that obtained in the direct calculation
of the degree distribution.

The fractality behavior can be also easy to see by tiling the system
using a renormalization procedure as follows. we can `zoom out'
(i.e. renormalize) the network by replacing each connected cluster
of bonds on the right of Fig.~\ref{fig1} by a single `super'-link,
in a way that reverses the process of network growth, see
Fig.~\ref{fig1}. This has the effect of rescaling diameter, it
reduces the diameter by a factor of $2$ by carrying a cluster of
bonds with diameter 2 into a single `super'-link with unit length.
At the same time, the number nodes of in the rescaled network
decreases by a factor $=4$. Thus, the fractal dimension is $d_B
=\frac{\ln 4}{\ln 2}=2$.

\section{SIR model on the networks}
As demonstrated in the preceding section, the networks exhibit
simultaneously many interesting properties: scale-free phenomenon,
tunable clustering, ``large-world" behavior, and fractality. To the
best of our knowledge, these features have not been reported in
previous network models. Hence, it is of scientific interest and
great significance to investigate dynamic processes taking place
upon the model to inspect the effect of network topologies on the
dynamics. Next we will focus on studying the SIR model of epidemics,
which is one of the most important disease models and has attracted
much attention within physical society, see~\cite{BaBaPaVe05} and
references therein.

The standard SIR model~\cite{Ba75,He00} assumes that each individual
can be in any of three exclusive states: susceptible (S), infected
(I), or removed (R). The dynamics of disease transmission on a
network can be described as follows. Each node of the network stands
for an individual and each link represents the connection along
which the individuals interact and the epidemic can be transmitted.
At each time step, an infected node transmits infection to each of
its neighbors independently with probability $\lambda$; at the same
time the infected node itself becomes removed with a constant
probability 1, whereupon it can not catch the infection again, and
thus will not infect its neighbors any longer.

The above-described SIR model for disease dynamics is equivalent to
a bond percolation process with bond occupation probability equal to
the infection rate $\lambda$~\cite{Gr82,Ne02}. The connected
clusters of nodes in the bond percolation process then correspond to
the groups of individuals that would be infected by a disease
outbreak starting with any individual within the cluster it belongs
to. It was shown that in scale-free networks such as
Barab\'asi-Albert (BA) network, for any vanishingly small $\lambda$,
there exists an extensive spanning cluster or ``giant component'',
implying that scale-free networks demonstrate an absence of epidemic
thresholds. However, we will present that for our networks, epidemic
thresholds are present, the values of which depend on the parameter
$q$.

\begin{figure}[h]
\centering\includegraphics[width=0.6\textwidth]{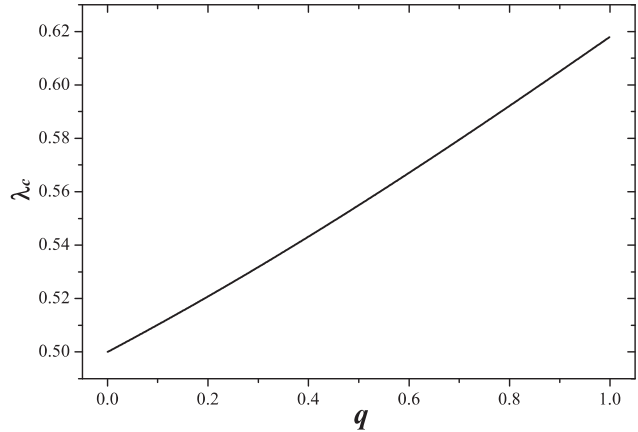} \caption{
The dependence relation of percolation threshold $\lambda_c$ on the
parameter $q$. }\label{fig4}
\end{figure}

According to the mapping between the SIR dynamics and the bond
percolation problem, the epidemic threshold corresponds precisely to
the percolation threshold. In our case, the recursive construction
of the networks make it possible to study the percolation problem by
using the real-space renormalization group
technique~\cite{MiEk7576,Ka76,Do03} to obtain an exact solution for
the percolation (epidemic) thresholds. Here we are interested in
only the percolation phase transition point, excluding the size of
giant cluster. Let us describe the procedure in application to the
considered network. Supposing that the network growth stops at a
time step $t\rightarrow \infty$, then we spoil the network in the
following way: for an arbitrary present link in the undamaged
network, it is retained in the damaged network with probability
$\lambda$. Then we invert the transformation shown in Fig. 1 and
define $n=t- \tau$ for this inverted transformation, which is in
fact a decimation procedure~\cite{Do03}. Further, we introduce the
probability $\lambda_n$ that if two nodes are connected in the
undamaged network at $\tau =t-n$, then at the $n$th step of the
decimation for the damaged network, there exists a path between
these vertices. Here, $\lambda_0= \lambda$. According to the network
construction and the analysis in~\cite{Do03}, it is easy to obtain
the following recursive relation for $\lambda_n$:
\begin{eqnarray}\label{lambda01}
\lambda_{n+1}=&\quad& q\left(\lambda_n^2+\lambda_n^2-\lambda_n^2\times\lambda_n^2\right)\nonumber\\
&+&(1-q)\Big[\lambda_n^5+5\lambda_n^4(1-\lambda_n)\nonumber\\
&+&8\lambda_n^3(1-\lambda_n)^2+2\lambda_n^2(1-\lambda_n)^3\Big].
\end{eqnarray}
Equation~(\ref{lambda01}) has five roots (i.e., fixed points), among
which two are invalid: one is greater than 1, the other is less than
0. The other three fixed points are as follows: two stable fixed
points at $\lambda=0$ and $\lambda=1$, and an unstable fixed point
at $\lambda_c$ that is the percolation threshold. We omit the
expression of $\lambda_c$ as a function of $q$, because it is very
lengthy. We show the dependence of $\lambda_c$ on $q$ in
Fig.~\ref{fig4}, which indicates that the threshold $\lambda_c$
increases almost linearly as $q$ increases. When $q$ grows from 0 to
1, $\lambda_c$ increases from 0.5 to $\frac{\sqrt{5}-1}{2}\approx
0.618$.

Therefore, there exists a percolation threshold $\lambda_c$ such
that for $\lambda >\lambda_c$ a giant component appears, for
$\lambda <\lambda_c$ there are only small clusters. This means that
for the SIR model the epidemic prevalence undergoes a phase
transition at a nonzero epidemic threshold $\lambda_c$. If the
infection rate $\lambda >\lambda_c$, the disease spreads and becomes
persistent in time; otherwise, the infection dies out gradually. The
existence of epidemic thresholds in our networks is in sharp
contrast with the null threshold found in a wide range of stochastic
scale-free networks of the Barab\'asi-Albert (BA)
type~\cite{PaVe01a,PaVe01b,MoPaVe02}.

Why are the present scale-free networks not prone to disease
propagation as previously studied uncorrelated BA type scale-free
networks? We argue that the presence of finite epidemic thresholds
in our networks lies with their two-dimensional fractal structure
with diameter increasing as a square power of network order, a
property analogous to that of two-dimensional regular
lattice~\cite{Ne00}. The `large-world' feature stops the diffusion
of diseases, and makes the behaviors of disease spreading in our
networks similar to those of regular lattices. Thus, the fractal
topology provides protection against disease spreading.

\section{Conclusions}

In the present work, we have introduced a new model for fractal
networks and provided a detailed analysis of the structural
properties, which are related to the model parameter $q$. The model
exhibits a rich topological behavior. The degree distribution is
power-law with the degree exponent asymptotically approaching 3 for
large network order. The clustering coefficient is changeable, which
can be systematically tunable in a large range by altering $q$.
Particularly, the networks are topologically fractal with a fractal
dimension of 2 for all $q$. Along with the fractality, the networks
display `large-world' phenomenon, their average path length
increases approximatively as a square power of the number of nodes.

We have also investigated the effects of the particular topological
characteristics on the SIR model for disease spreading dynamics.
Strikingly, We found that epidemic thresholds are recovered for all
networks regardless the value of $q$, which is in contrast to the
conventional wisdom that being prone to disease propagation is an
intrinsic nature of scale-free networks. We concluded that the
dominant factor suppressing epidemic spreading is the fractal
structure accompanied by a `large-world' behavior. The peculiar
structural properties and epidemic dynamics make our networks unique
within the category of scale-free networks. Our study is helpful for
designing real-life networked systems robust to epidemic outbreaks,
and for better understanding of the effluences of structure on the
propagation dynamics.

\section*{Acknowledgment}
We thank Yichao Zhang for preparing this manuscript. This research
was supported by the National Basic Research Program of China under
grant No. 2007CB310806, the National Natural Science Foundation of
China under Grant Nos. 60496327, 60573183, 90612007, 60773123, and
60704044, the Shanghai Natural Science Foundation under Grant No.
06ZR14013, the China Postdoctoral Science Foundation funded project
under Grant No. 20060400162, the Program for New Century Excellent
Talents in University of China (NCET-06-0376), and the Huawei
Foundation of Science and Technology (YJCB2007031IN).

\end{document}